# Full state revivals in higher dimensional quantum walks

Mahesh N. Jayakody and Asiri Nanayakkara*
*National Institute of Fundamental Studies,*
*Hanthana Road, Kandy, Sri Lanka*

## Abstract

Full state revivals in a quantum walk can be viewed as returning of the walker to the initial quantum state in a periodic fashion during the propagation of the walk. In this paper we show that for any given number of spatial dimensions, a coin operator can be constructed to generate a quantum walk having full revivals with any desired period. From the point of view of quantum computation and simulations, these coin operators can be useful in implementing quantum walks which oscillate between any two states with a finite periodicity.

## 1. Introduction

Quantum walks have become a fruitful testing ground in various areas of science during the past few decades. Principally they contribute to theoretical and practical improvements in quantum algorithms [1-4] and quantum computing [5, 6]. In addition, quantum walks have been used to model transport in biological systems [7-9] and other physical phenomena such as Anderson localization [10-14] and topological phases [15, 16].

Occurrence of full-revivals in the evolution of quantum walks is an interesting feature which has no direct analogue in classical random walks. Reappearance of an arbitrary state; (both position and coin states) with a finite time gap is recognized as full-revivals in a quantum walk. With regard to the above definition, a full state revival has no meaning in the context of classical random walks. Instead, one can ask about the probability that a walker engaged in a classical random walk returns to the starting point during the evolution. This is termed as the recurrence probability which is characterized by the P$ó$lya number. An extensive work related to recurrence probability has been carried out in the context of quantum walks by defining a quantum version of classical P$ó$lya number [17-20].

In concurrent studies, occurrence of full states revivals, which is unique to quantum walks has also been investigated extensively. Ref [21] shows that four-state Grover walk results in 2 step full-revival and further, it has been proved that periods more than 2 steps cannot be achieved for a four-state quantum walk in general. On the contrary, in our study we have observed a 4 step full-revivals in a four-state quantum walk. The study conducted in [22] is concerned with the conditions under which a quantum walker on cyclic path will returns to its initial quantum state in $n$ steps. In [23] it has been shown that Hardmard Walk in a cycle can have only 2, 4 and 8 steps periods for full-state revivals. Ref [24] provides an experimental realization of two periodic revivals in a single-photon one dimensional quantum walk with a linearly ramped time-dependent coin flip. By employing a spatial dependent coin rather than a time dependent coin, a theoretical explanation for the aforementioned experiment is given in [25].

In our approach we define a class of evolution operators that can generate quantum walks with full-state revivals. No time dependent coins or shift operators are used in this method. Periodicity in full-revivals solely depends upon the coin dimension. Moreover, we show that for a given number of spatial dimensions, one can construct a quantum walk having any period. First we prove our result for the quantum walk on a line with $n$ coin states and later generalize it into higher spatial dimensions.

## 2. Quantum walk on a line

First let us recall the standard model of the quantum walk on a line that comprises a two-state coin and a walker. Evolution of the coin-walker system is governed by a unitary operator $U$ defined on a tensor product of two Hilbert spaces, $H_c \otimes H_x$ which are spanned by the coin basis $\{|c\rangle\}_{c \in \{0,1\}}$ and the position basis $\{|x\rangle\}_{x \in Z}$. Single-step progression of the system is a sequential process in which the coin is tossed at first (transforming the coin state) and then the walker is moved either to the left or right conditional upon the outcome of the coin. Unitary operator that corresponds to a single-step evolution of the system is given by

$$U = S \cdot (\mathbb{I} \otimes C) \tag{1}$$

where $S$ and $C$ are Shift and Coin operators respectively. In this study we stick ourselves to the conventional shift operator $S$ given by

$$S = \sum_{x,c=0,1} |c\rangle\langle c| \otimes |x + (-1)^{c+1}\rangle\langle x| \tag{2}$$

General form of a unitary coin operator [26] that governs the 1D quantum walks can be written as

$$C = \begin{pmatrix} \cos(\theta) & e^{i\phi_1}\sin(\theta) \\ e^{i\phi_2}\sin(\theta) & -e^{i(\phi_1+\phi_2)}\cos(\theta) \end{pmatrix} \tag{3}$$

where $\theta \in [0, 2\pi)$, and $\phi_1, \phi_2 \in [0, \pi)$. The general state vector of the coin-walker system is written as

$$|\psi(x,t)\rangle = \sum_x a_x(t)|0\rangle|x\rangle + b_x(t)|1\rangle|x\rangle \tag{4}$$

where $\sum_x |a_x(t)|^2 + |b_x(t)|^2 = 1$ and $t$ is the time step. Let the initial state of the system be $|\psi_0\rangle$. Thus the final state of the system after $t$ steps is $|\psi(x,t)\rangle = U^t|\psi_0\rangle$.

In general, coin Hilbert space of a quantum walk which has two coin states, is spanned by a basis set with two distinct elements. Since the Hilbert spaces of same dimensions are isomorphic, we can easily find a transformation from coin Hilbert space to a matrix Hilbert space. In other words we can find a one to one correspondence with coin states and n×1 column vectors where n is the dimension of the Hilbert space. Such a correspondence allows us to represent all the coin states in terms of n×1 column vectors and all the operators in coin Hilbert space as n×n matrixes. In the same fashion, position Hilbert space can also be represented in the matrix form. However, the position space is spanned by a basis set with infinite number of elements and hence, the matrix representation contains infinite number of matrix elements. For a finite number of steps, position states can be represented as column vector with finite number of elements. Yet, as the time evolves the dimension of the column vectors that represent the positon tends to increase. Consequently, the unitary operator that governs the quantum walk cannot be represented as a block matrix in a convenient way for long time limits. In addressing this issue, momentum representation of the position space provides a viable solution by embedding all the information related to shift operation into coin operation. Propagator that governs a quantum walk on a line with two coin states can be represented in momentum space as follows

$$U_k = C_k \qquad (5)$$

where $C_k = (e^{-ik}P_R + e^{ik}P_L)C$. Note that $P_R$ and $P_L$ are two orthogonal projectors on coin Hilbert space where $P_R + P_L = \mathbb{I}$ and $C$ is the usual coin operator [27]. Further, note that the time evolution in the momentum representation is governed by the propagator $C_k$. Unlike in positon representation the dimension of the propagator $C_k$ does not increase with time. Hence, the behavior of the quantum walk can be determined by diagonalizing the unitary operator $C_k$. In addition, the quantum state in the position representation at any time $t$ can be determined simply by taking the inverse Fourier transform of the corresponding state in the momentum representation after applying the unitary operator $C_k$ on the initial state $t$ number of times.

It is possible to define a quantum walk on a line by incorporating more than two coin states. Let us consider a set of orthonormal vectors $\{|\phi_r\rangle\}_{r=1}^n$. We can define a quantum walk on a line by assigning a separate shifting rule for each state $|\phi_r\rangle$. Coin space of such a quantum walk is spanned by the orthonormal set $\{|\phi_r\rangle\}_{r=1}^n$ and the coin operator can be represented as a $n \times n$ unitary matrix with respect to the spanning set. In general, any operator of the following form represents a coin operator for a quantum walk on a line with $n$-coin states.

$$C^{(n)} = \sum_{r,s=1}^n \lambda_{r,s}|\phi_r\rangle\langle\phi_s| \qquad (6)$$

Since $C^{(n)}$ satisfies the unitary condition, $\lambda_{r,v}\lambda_{w,s}^* = \delta_{r,w}\delta_{v,s}$. Let $\{|\phi_i\rangle\langle\phi_i|\}_{i=1}^n$ be a set of orthogonal projectors defined on the coin Hilbert space. Thus in the momentum representation we can define the propagator that governs the evolution of a quantum walk on a line with $n$-coin states as

$$U_k^{(n)} = \left(\sum_{w=1}^n e^{ia_wk}|\phi_w\rangle\langle\phi_w|\right)\left(\sum_{r,s=1}^n \lambda_{r,s}|\phi_r\rangle\langle\phi_s|\right)$$

$$U_k^{(n)} = \sum_{r,s=1}^n \lambda_{r,s}e^{ia_rk}|\phi_r\rangle\langle\phi_s| \qquad (7)$$

where $\lambda_{r,v}\lambda_{w,s}^* = \delta_{r,w}\delta_{v,s}$, $k \in [-\pi, \pi)$ and for each $w \in \{1, ..., n\}$, $a_w$ is an integer. Usual choice for integers $a_w$ is defined as follows:

If $n$ is even: $a_w \in \left\{\left(-\frac{n}{2}\right), \left(-\frac{n}{2}+1\right), \ldots, -2, -1, 1, 2, \ldots, \left(\frac{n}{2}-1\right), \left(\frac{n}{2}\right)\right\}$

If $n$ is odd: $a_w \in \left\{\left(-\frac{n-1}{2}\right), \left(-\frac{n-1}{2}+1\right), \ldots, -2, -1, 0, 1, 2, \ldots, \left(\frac{n-1}{2}-1\right), \left(\frac{n-1}{2}\right)\right\}$

E.g.: when $n = 2$; possible shifts are $\pm 1$ while when $n = 3$; possible shifts are zero and $\pm 1$

## 3. Periodic unitary evolutions in Hilbert spaces

In a finite dimensional Hilbert space, the necessary condition for a periodic unitary evolution, governed by a unitary operator $W$ is $W^T = \mathbb{I}$, for some $T \in \mathbb{N}$ where $T$ is the period. In this section we show that for a given set of orthonormal vectors in a finite dimensional Hilbert space, one can always construct a unitary operator that can generate a periodic evolution.

**Proposition 1:**

Let $\{|\phi_1\rangle, \ldots, |\phi_n\rangle\}$ be a set of orthonormal vectors in a Hilbert space $H$ where $n, \in \mathbb{N}$ and $dim(H) \geq n > 1$. Define an operator $W$ of the form

$$W = \sum_{j=2}^{n} \lambda_j |\phi_j\rangle\langle\phi_{j-1}| + \lambda_1 |\phi_1\rangle\langle\phi_n| \tag{8}$$

where for each $j \in \{1, \ldots, n\}$ $\lambda_j \in \mathbb{C}$ such that $\prod_{j=1}^{n} \lambda_j = 1$. Then $W^n = \mathbb{I}$ where $\mathbb{I}$ is the identity operator of the subspace of $H$ spanned by $\{|\phi_1\rangle, \ldots, |\phi_n\rangle\}$.

**Proof:**

The $m^{th}$ power of $W$ where $1 < m < n$ can be written as

$$W^m = \sum_{k=m+1}^{n}\left(\prod_{l=k-(m-1)}^{k} \lambda_l\right)|\phi_k\rangle\langle\phi_{k-m}| + \sum_{j=1}^{m-1}\left(\prod_{u=1}^{j} \lambda_u\right)\left(\prod_{v=0}^{m-j-1} \lambda_{n-v}\right)|\phi_j\rangle\langle\phi_{n-(m-j)}| + \left(\prod_{u=1}^{m} \lambda_u\right)|\phi_m\rangle\langle\phi_n| \tag{9}$$

Using the method of induction it can be proved that (9) is valid for each $m$ where $1 < m < n$. By substituting $m = n - 1$ we get:

$$W^{n-1} = \left(\prod_{l=2}^{n} \lambda_l\right)|\phi_n\rangle\langle\phi_1| + \left(\prod_{l=1}^{n-1} \lambda_l\right)|\phi_{n-1}\rangle\langle\phi_n| + \sum_{j=1}^{n-2}\left(\prod_{u=1}^{j} \lambda_u\right)\left(\prod_{v=0}^{n-j-2} \lambda_{n-v}\right)|\phi_j\rangle\langle\phi_{j+1}| \tag{10}$$

Multiplying (8) by (10), an expression for the operator $W^n$ can be written as follows:

$$W^n = \left(\prod_{l=1}^{n} \lambda_l\right)|\phi_n\rangle\langle\phi_n| + \left(\prod_{l=1}^{n} \lambda_l\right)|\phi_{n-1}\rangle\langle\phi_{n-1}| + \sum_{j=1}^{n-2}\left(\prod_{u=1}^{j} \lambda_u\right)\left(\prod_{v=0}^{n-j-2} \lambda_{n-v}\right)\lambda_{j+1}|\phi_j\rangle\langle\phi_j| \tag{11}$$

Note that for each $j$ the following relationship holds:

$$\left(\prod_{u=1}^{j} \lambda_u\right)\left(\prod_{v=0}^{n-j-2} \lambda_{n-v}\right)\lambda_{j+1} = \left(\prod_{l=1}^{n} \lambda_l\right) \tag{12}$$

Thus we can write $W^n = \left(\prod_{j=1}^{n} \lambda_j\right)\mathbb{I}$. Since $\left(\prod_{j=1}^{n} \lambda_j\right) = 1$ and we have $W^n = \mathbb{I}$. This completes the proof. Note that for each $j$ we can always choose $\lambda_j = e^{i\theta_j}$ where $\theta_j \in (-\pi, \pi]$ such that $W$ is unitary.

**Lemma:**
Let $\{|\phi_1\rangle, \ldots, |\phi_r\rangle, \ldots, |\phi_n\rangle\}$ be a set of orthonormal vectors in a Hilbert space $H$ where $n, r \in \mathbb{N}$ and $dim(H) \geq n \geq r > 1$. Define an operator $W$ of the form

$$W = \sum_{i,j=1}^{n} \mu_{i,j} |\phi_i\rangle\langle\phi_j|$$

where

$$\mu_{i,j} = \begin{cases} \lambda_i(\delta_{i,j+1} + \delta_{i,1}\delta_{r,j}), & i \leq r \\ \delta_{i,j}, & i > r \end{cases} \quad (13)$$

and for each $i \in \{1, \ldots, r\}$, $\prod_{i=1}^{r} \lambda_i = 1$ where $\lambda_i \in \mathbb{C}$. Then $(W)^r = \mathbb{I}$ where $\mathbb{I}$ is the identity operator of the subspace of $H$ spanned by $\{|\phi_1\rangle, \ldots, |\phi_n\rangle\}$.

**Proof:**
The equation (13) is written as

$$W = \sum_{j=1}^{n}\sum_{i=1}^{r} \lambda_i(\delta_{i,j+1} + \delta_{i,1}\delta_{r,j})|\phi_i\rangle\langle\phi_j| + \sum_{i=r+1}^{n} |\phi_i\rangle\langle\phi_i| \quad (14)$$

Define $W' = \sum_{j=1}^{n}\sum_{i=1}^{r} \lambda_i(\delta_{i,j+1} + \delta_{i,1}\delta_{r,j})|\phi_i\rangle\langle\phi_j|$. By re-indexing $W'$ we can write $W' = \sum_{i=2}^{r} \lambda_i|\phi_i\rangle\langle\phi_{i-1}| + \lambda_1|\phi_1\rangle\langle\phi_r|$. Since all the vectors in $\{|\phi_1\rangle, \ldots, |\phi_r\rangle, \ldots, |\phi_n\rangle\}$ are orthonormal, $r^{th}$ power of $W$ can be written as $(W)^r = (W')^r + \sum_{i=r+1}^{n}|\phi_i\rangle\langle\phi_i|$. From the proposition 1, we have $(W')^r = \sum_{i=1}^{r}|\phi_i\rangle\langle\phi_i|$. Thus $(W)^r = \mathbb{I}$ where $\mathbb{I}$ is the identity operator of the subspace of $H$ spanned by $\{|\phi_1\rangle, \ldots, |\phi_n\rangle\}$. This completes the proof. Note that for each $j \in \{1, \ldots, r\}$ we can always choose $\lambda_j = e^{i\theta_j}$ where $\theta_j \in (-\pi, \pi]$ such that $W$ is unitary.

## 4. Full revivals in Quantum walk on a line

Before moving in to higher spatial dimensions let us consider a case related to full-state reveals in quantum walk on a line. Without using the usual case of two coin sates, here we consider a quantum walk on a line with $n$-coin states. Being consistent with the known facts, in order to exhibit full-quantum states revivals during the propagation, the propagator of a quantum walk on a line with $n$-coin states given in (7) must satisfy the condition $\left(U_k^{(n)}\right)^m = \mathbb{I}$ where $m \in \mathbb{N}$. It is fairly difficult to determine an explicit form for the $n \times n$ unitary matrix $U_k^{(n)}$ while satisfying the aforementioned condition for revivals. In handling that problem one can impose conditions on $U_k^{(n)}$ in such a way that each of the eigenvalue $\lambda$ of $U_k^{(n)}$ satisfies the property $\lambda^m = 1$. Then by diogaonalizing the matrix $U_k^{(n)}$ we can easily show that the $m^{th}$ power of $U_k^{(n)}$ is equal to the unit operator. This approach is followed in [22] for a general quantum walk on a cycle. Intuitively, it can be stated that the recursive nature embedded in the cyclic walks makes it easy to find such conditions to generate revivals. However, determining revival condition for a quantum walk on non-cycles might be rather challenging. As a solution to this problem, we present a different approach by defining a specific form for $U_k^{(n)}$ in such a way that it always satisfies the condition for revival. We show that any propagator in this form can exhibit full quantum state revivals in quantum walks on non-cyclic paths.

The operator given in (8) can be used to define a set of propagators that can produce quantum walks with full state revivals having periodicity equal to the dimension of coin space. Let $\{|\phi_1\rangle, ..., |\phi_n\rangle\}$ be the basis set which spans a given coin Hilbert space where $n \in \mathbb{N}$ and $n > 1$. Now let us define a new coin operator of the form

$$W^{(n)} = \sum_{j=2}^{n} e^{i\theta_j}|\phi_j\rangle\langle\phi_{j-1}| + e^{i\theta_1}|\phi_1\rangle\langle\phi_n| \tag{15}$$

where for each $j \in \{1, ..., n\}$, $\theta_j \in (-\pi, \pi]$ and $\sum_{j=1}^{n} \theta_j = 2g\pi$, $g \in \mathbb{Z}$. Note that $W^{(n)}(W^{(n)})^\dagger = (W^{(n)})^\dagger W^{(n)} = \mathbb{I}$. Let $\{|\phi_i\rangle\langle\phi_i|\}_{i=1}^{n}$ be a set of orthogonal projectors defined on coin Hilbert space. Now we construct an operator of the form

$$D^{(n)} = \sum_{j=1}^{n} e^{ia_j k}|\phi_j\rangle\langle\phi_j| \tag{16}$$

where $k \in [-\pi, \pi)$ and for each $j \in \{1, ..., n\}$, $a_j$ is an integer such that $\sum_{j=1}^{n} a_j = 0$. It is obvious that $D^{(n)}$ is unitary. By combining (15) and (16) we can define a propagator $V_k^{(n)}$ in the momentum space which governs a quantum walk on a line with $n$-coin states as;

$$V_k^{(n)} = D^{(n)} W^{(n)} = \sum_{j=2}^{n} e^{ia_j k} e^{i\theta_j}|\phi_j\rangle\langle\phi_{j-1}| + e^{ia_1 k} e^{i\theta_1}|\phi_1\rangle\langle\phi_n| \tag{17}$$

Matrix form of $V_k^{(n)}$ with respective to orthonormal basis $\{|\phi_1\rangle, ..., |\phi_n\rangle\}$ is given by

$$V_k^{(n)} = \begin{pmatrix} 0 & 0 & 0 & \cdots & 0 & 0 & e^{ia_1 k} e^{i\theta_1} \\ e^{ia_2 k} e^{i\theta_2} & 0 & 0 & \cdots & 0 & 0 & 0 \\ 0 & e^{ia_3 k} e^{i\theta_3} & 0 & \cdots & 0 & 0 & 0 \\ \vdots & \vdots & \vdots & \ddots & \vdots & \vdots & \vdots \\ 0 & 0 & 0 & \cdots & 0 & 0 & 0 \\ 0 & 0 & 0 & \cdots & 0 & 0 & 0 \\ 0 & 0 & 0 & \cdots & 0 & e^{ia_n k} e^{i\theta_n} & 0 \end{pmatrix} \tag{18}$$

Then operator $V_k^{(n)}$ takes the form of the operator $W$ given in (8). Note that $\sum_{j=1}^{n} \theta_j = 2g\pi$, $g \in \mathbb{Z}$ and $\sum_{j=1}^{n} a_j = 0$. Hence we have

$$\left(V_k^{(n)}\right)^n = \mathbb{I} \tag{19}$$

Thus indeed the propagator $V_k^{(n)}$ can give rise to full quantum state revivals. Therefore $V_k^{(n)}$ represents a quantum walk on line with $n$-coin states, governed by the coin $W^{(n)}$, that can exhibit full state revivals with a periodicity of $n$ time steps. Periodicity of revivals solely depends upon the dimension of the coin space. As a result, we can define different propagators simply by adding new coin degrees of freedom and can generate full revivals with different periods. For example, we can define a propagator for a quantum walk on a line with two coin states and it has a period of 2 steps. By adding a new state orthogonal to the existing coin states we can define a new propagator for a quantum walk on a line with three coin states. Thus latter quantum walk has a period of 3 steps. Let us see some examples. Consider a quantum walk on a line with two orthonormal coin states $\{|c\rangle\}_{c=1}^{2}$. Define a coin operator of the form $C^{(2)} = e^{\frac{4\pi i}{3}}|1\rangle\langle 2| + e^{\frac{2\pi i}{3}}|2\rangle\langle 1|$. Note that $C^{(2)}$ has the form of one-dimensional Grover coin. Let the shift operator be $S^{(2)} = \sum_{x,c=1,2}|c\rangle\langle c| \otimes |x+(-1)^c\rangle\langle x|$. Table 1 shows the first three states of the quantum walk which is governed by the unitary operator $U^{(2)} = S^{(2)}(\mathbb{I}\otimes C^{(2)})$

| Time | State |
|---|---|
| t=0 (initial state) | $\frac{1}{\sqrt{2}}(|1_c, 1_x\rangle + |2_c, 1_x\rangle)$ |
| t=1 | $\frac{1}{\sqrt{2}}e^{\frac{-2\pi i}{3}}|1_c, 0_x\rangle + \frac{1}{\sqrt{2}}e^{\frac{2\pi i}{3}}|2_c, 2_x\rangle$ |
| t=2 | $\frac{1}{\sqrt{2}}(|1_c, 1_x\rangle + |2_c, 1_x\rangle)$ |

Table1: Full revivals of quantum states with a period of 2 steps ($U^{(2)} = S^{(2)}(\mathbb{I}\otimes C^{(2)})$)

Now consider a quantum walk on a line with three orthonormal coin states $\{|c\rangle\}_{c=1}^{3}$. Define a coin operator of the form $C^{(3)} = |1\rangle\langle 3| + e^{\frac{2\pi i}{3}}|2\rangle\langle 1| + e^{\frac{4\pi i}{3}}|3\rangle\langle 2|$. Let the shift operator be $S^{(3)} = |1\rangle\langle 1| \otimes |x-5\rangle\langle x| + |2\rangle\langle 2| \otimes |x+3\rangle\langle x| + |3\rangle\langle 3| \otimes |x+2\rangle\langle x|$. Note that the total shift (-5, 3, and 2) is chosen to be zero. Table 2 shows the first four states of the quantum walk which is governed by the unitary operator $U^{(3)} = S^{(3)}(\mathbb{I}\otimes C^{(3)})$.

| Time | State |
|---|---|
| t=0 (initial state) | $\frac{1}{\sqrt{3}}(|1_c, 3_x\rangle + |2_c, 2_x\rangle + |3_c, 1_x\rangle)$ |
| t=1 | $\frac{1}{\sqrt{3}}|1_c, -4_x\rangle + \frac{1}{\sqrt{3}}e^{\frac{2\pi i}{3}}|2_c, 6_x\rangle + \frac{1}{\sqrt{3}}e^{\frac{-2\pi i}{3}}|3_c, 4_x\rangle$ |
| t=2 | $\frac{1}{\sqrt{3}}e^{\frac{-2\pi i}{3}}|1_c, -1_x\rangle + \frac{1}{\sqrt{3}}e^{\frac{2\pi i}{3}}|2_c, -1_x\rangle + \frac{1}{\sqrt{3}}|3_c, 8_x\rangle$ |
| t=3 | $\frac{1}{\sqrt{3}}(|1_c, 3_x\rangle + |2_c, 2_x\rangle + |3_c, 1_x\rangle)$ |

Table2: Full revivals of quantum states with a period of 3 steps ($U^{(3)} = S^{(3)}(\mathbb{I}\otimes C^{(3)})$)

Properties of a given quantum walk are attributed by the eigen spectrum of the propagator that governs the walk. Hence it is worth analyzing the spectrum of the class of operators given in (17). Since $V_k^{(n)}$ is unitary, all the eigen values must lie on a unite circle in the complex plane. Most of the elements of the matrix given in (18) are zeros and hence the determinant of $\left(V_k^{(n)} - \lambda \mathbb{I}\right)$ can easily be calculated. The expression for the determinant of $\left(V_k^{(n)} - \lambda \mathbb{I}\right)$ is given by;

$$\det\left(V_k^{(n)} - \lambda \mathbb{I}\right) = \lambda^n - \left(\prod_{j=1}^{n} e^{ia_j k} e^{i\theta_j}\right) \quad (20)$$

Since $\prod_{j=1}^{n} e^{ia_j k} e^{i\theta_j} = 1$ we can determine the eigenvalues of $V_k^{(n)}$ by solving $\lambda^n - 1 = 0$. In other words, $n^{th}$ roots of unity become the eigenvalues of the operator $V_k^{(n)}$. The significant feature of this is that all of the eigen values of $V_k^{(n)}$ are independent of $k$. Hence the corresponding propagator in the positon representation has a point spectrum with the same eigen values. In addition, the corresponding eigen vectors characterize the stationary states in the position space.

It is worth mentioning that by using the operator in (13), one can define a coin for a quantum walk on a line having a period less than the dimension of the coin space. For this purpose, zero shifting rule (particle is allowed stay at the same position) must be assigned to certain coin degrees of freedom ($\{|\phi_i\rangle\}_{i=r+1}^{n}$). Even though such a walk exhibits full-state revivals it will always have some localized position states.

## 5. Full revivals in higher spatial dimensions

An interesting question that one could ask regarding the full-state revivals is, for a given $n$-spatial dimensions whether we can construct a coin and a shift operator that can generate a quantum walk having a revival of period $T$. The method given below provides one possible solution to this question. First we show the existence of such a solution and later give a concrete example.

**Proposition 2:**
Given a spatial dimension $n \in \mathbb{N}\backslash\{0\}$ and a positive integer $T$ there exist a quantum coin which drives a quantum walk that exhibit full-state revivals with a period $T$.

**Proof:**
Let $B = \{|\phi_1\rangle, ..., |\phi_T\rangle\}$ be an orthonormal basis for a $T$-dimensional coin Hilbert space $H_T$. Define a coin operator of the form $W^{(T)} = \sum_{j=2}^{T} e^{i\theta_j}|\phi_j\rangle\langle\phi_{j-1}| + e^{i\theta_1}|\phi_1\rangle\langle\phi_T|$ where for each $j \in \{1, ..., T\}$ we have $\theta_j \in (-\pi, \pi]$ such that $\sum_{j=1}^{T} \theta_j = 2g\pi$, $g \in \mathbb{Z}$. It can be easily proved that $W^{(T)}$ is unitary. Define a subset $B_r$ of $B$ for each $r^{th}$ spatial dimension where $r \in \{1, ..., n\}$ such that $|B_r| \geq 2$. Now let us define the shifting rule in each spatial dimension. Consider a function of the form

$$a_{r,j} = \begin{cases} a_{r,j} \in \mathbb{Z}, & \text{if } |\phi_j\rangle \in B_r \\ 0, & \text{otherwise} \end{cases} \quad (21)$$

where $j \in \{1, ..., T\}$ such that $\sum_{j=1}^{T} a_{r,j} = 0$. Define an operator $D^{(T)}$ such that

$$D^{(T)} = \sum_{j=1}^{T} e^{-i(\sum_{r=1}^{n} a_{r,j}k_r)}|\phi_j\rangle\langle\phi_j| \qquad (22)$$

where for each $r \in \{1, ..., n\}$, $k_r \in [-\pi, \pi)$. Note that $D^{(T)}$ is unitary. By combing $W^{(T)}$ and $D^{(T)}$ we can construct a propagator $V_k^{(T)}$ in the momentum space which governs a quantum walk on $n$- spatial dimensional lattice as;

$$V_k^{(T)} = D^{(T)} W^{(T)} = \sum_{j=2}^{T} e^{-i(\sum_{r=1}^{n} a_{r,j}k_r)} e^{i\theta_j}|\phi_j\rangle\langle\phi_{j-1}| + e^{-i(\sum_{r=1}^{n} a_{r,1}k_r)} e^{i\theta_1}|\phi_1\rangle\langle\phi_T| \qquad (23)$$

Note that $V_k^{(T)}$ takes the form of the operator given in (17). Thus we can write

$$\left(V_k^{(T)}\right)^T = \prod_{j=1}^{T}\left(e^{-i(\sum_{r=1}^{n} a_{r,j}k_r)} e^{i\theta_j}\right)\mathbb{I} \qquad (24)$$

Note that $\prod_{j=1}^{T}\left(e^{-i(\sum_{r=1}^{n} a_{r,j}k_r)} e^{i\theta_j}\right) = e^{-i\left(\sum_{r=1}^{n}\left(k_r \sum_{j=1}^{T} a_{r,j}\right)\right)} e^{i\sum_{j=1}^{T} \theta_j} = 1$. Hence we have

$$\left(V_k^{(T)}\right)^T = \mathbb{I} \qquad (25)$$

This completes the proof. For further clarification let us consider an example. Suppose $n = 2$ and $T = 3$. Consider an orthonormal coin basis $\{|c\rangle\}_{c=1}^{3}$. Define a coin operator of the form $C' = e^{\frac{2\pi i}{3}}|2\rangle\langle 1| + e^{\frac{4\pi i}{3}}|3\rangle\langle 2| + |1\rangle\langle 3|$. Let the shift operator be $S'=|1\rangle\langle 1| \otimes |x+1, y-1\rangle\langle x, y| + |2\rangle\langle 2| \otimes |x+1, y-1\rangle\langle x, y| + |3\rangle\langle 3| \otimes |x-2, y-2\rangle\langle x, y|$. Table 3 shows the first three states of the quantum walk which is governed by the unitary operator $U' = S'(\mathbb{I} \otimes C')$

| Time | State |
|---|---|
| t=0 (initial state) | $\frac{1}{\sqrt{3}}(|1_c, 0_x, 0_y\rangle + |2_c, 0_x, 0_y\rangle + |3_c, 0_x, 0_y\rangle)$ |
| t=1 | $\frac{1}{\sqrt{3}}|1_c, 1_x, -1_y\rangle + \frac{1}{\sqrt{3}}e^{\frac{2i\pi}{3}}|2_c, 1_x, -1_y\rangle + \frac{1}{\sqrt{3}}e^{-\frac{2i\pi}{3}}|3_c, -2_x, 2_y\rangle$ |
| t=2 | $\frac{1}{\sqrt{3}}e^{-\frac{2i\pi}{3}}|1_c, -1_x, 1_y\rangle + \frac{1}{\sqrt{3}}e^{\frac{2i\pi}{3}}|2_c, 2_x, -2_y\rangle + \frac{1}{\sqrt{3}}|3_c, -1_x, 1_y\rangle$ |
| t=3 | $\frac{1}{\sqrt{3}}(|1_c, 0_x, 0_y\rangle + |2_c, 0_x, 0_y\rangle + |3_c, 0_x, 0_y\rangle)$ |

Table 3: Full revivals of quantum states with a period of 3 steps ($U' = S'(\mathbb{I} \otimes C')$

# 6. Conclusion

In this paper we showed that for any given number of spatial dimensions we can construct a coin operator that generates a quantum walk having full revivals with any desired period. One of the possible applications of revival dynamics in quantum walks is generating oscillatory motions between two given states. From the point of view of quantum computation and simulations, the above mentioned coin operators can be useful in implementing quantum walks which toggle between given states with a finite periodicity.